\documentclass{article}

\usepackage[english]{babel}

\usepackage[letterpaper,top=2cm,bottom=2cm,left=3cm,right=3cm,marginparwidth=1.75cm]{geometry}

\usepackage{amsmath}
\usepackage{amssymb}
\usepackage{amsfonts}
\usepackage{amsthm}
\usepackage{graphicx}
\usepackage[colorlinks=true, allcolors=blue]{hyperref}
\newtheorem{remark}{Remark}

\begin{document}

\begin{center}
\large{\bf Singularity confinement and proliferation of tau functions for a general differential-difference Sawada-Kotera equation}
\end{center}
\vspace{0.8 cm}
\begin{center}
{\bf A. Marin (1, 2), A. S. Carstea (2)}\footnote{email: acarst@theor1.theory.nipne.ro, 
carstea@gmail.com}\\
\vspace{0.4 cm}
{\it (1) Department of Physics, University of Bucharest,}\\
{\it Magurele, P.O. Box MG11 Bucharest, ROMANIA}\\
\vspace{0.1 cm}
{\it (2) Department of Theoretical Physics, Institute of Physics and Nuclear
Engineering,}\\
{\it Magurele, P.O. Box MG6 Bucharest, ROMANIA}\\
\end{center}
\vspace{1 cm}

%\
%\maketitle
\begin{abstract}
 By blending Painlev\'e property with singularity confinement for a general arbitrary order Sawada-Kotera differential-difference equation, we find a proliferation of ``tau-functions'' (coming from confined patterns). However, only one of these function enters into the Hirota bilinear form (the others give multi-linear expressions) but it has specific relations with all others. We also discuss two modifications of the Sawada-Kotera equation. Fully discretizations and the express method for computing algebraic entropy are discussed.
\end{abstract}

\section{Introduction}
Singularity confinement is a very efficient tool in detecting possible integrable discrete systems. For finite dimensional case (mappings) it imposes that a finite number of iterations are needed for exiting singular behaviors and recovering the starting initial data. It was instrumental in finding discrete Painlev\'e equations \cite{alfred-prl}, \cite{basil-book} some years ago. Later on, Sakai \cite{sakai} realized that singularity confinement is intimately related to the classical desingularization (blowing-up/down) procedure in birational algebraic geometry and mappings are turned into regular automorphisms of rational/elliptic surfaces (or family of isomorphisms in the case of non-autonomous mappings related to singular fibers of an invariant elliptic fibration \cite{fane1}). However, singularity confinement is not sufficient for proving integrability. There are mappings which are confining but display chaotic behavior \cite{hv}.

Zero algebraic entropy or algebraic growth of the degree of iterates are considered sufficient for proving integrability \cite{basil-book}, \cite{jarmo-book}. However, very recently it was shown by Halburd \cite{halburd} that, from the structure of confining patterns, one can estimate the complexity growth and the value of the algebraic entropy. Based on this method, a simplified version (called {\it express method}) which allows the computation of algebraic entropy only, was developed \cite{mase-1}, \cite{mwrg}. Singularity confinement can also be applied in the case of infinite dimensional discrete systems (but for algebraic entropy one must be careful with respect to the initial data \cite{mase-2}).

An extremely important consequence of singularity confinement is the relation with Hirota bilinear formalism and tau-functions. The positions of tau-functions and Hirota substitutions are given by confining singularity patterns as well as by the affine Weyl groups associated to the resolution of singularities (and this was the first approach to the bilinear form of discrete Painlev\'e equations \cite{conte}, \cite{kny}; see also examples in \cite{alfred-fane1}, \cite{alfred-fane2} showing the connection with various singularity patterns).
For example, in the q-Painlev\'e VI equation the singularity patterns determine everything; the equation is nothing more than a way to represent different singularity patterns in terms of an entire function (the tau-function).

In this paper we intend to analyze the singularity confinement of some differential-difference systems, namely a class of Sawada-Kotera-type equations. Here we encounter a mixed situation. First of all, these equations are infinite dimensional and we cannot apply at all the machinery of desingularization by blowing-ups from algebraic geometry, which works only for finite dimensional case, and, secondly, here we have a continuous variable involved. Accordingly, the movable singularity (in the ``continuous'' part) is expressed as a Laurent series around it and, in turn, this series is iterated producing various singularity patterns. The study of singularities proved to be very useful in this case \cite{delay1},\cite{delay2} for analyzing delay-Painlev\'e equations \cite{delay3}\cite{alex} and integro-differential singular equations. In higher dimensional differential-difference case it is not clear how to rigorously define singularity confinement and we rely only on the number of fixed entries in the coordinates. These are instrumental in finding positions of tau functions. In \cite{fane}, for the case of Bogoyavlenski lattices, many {\it different} confining patterns corresponding to the same dependent variable and accordingly different representations in terms of tau functions were found. So we can speak about a {\it proliferation of tau-functions} corresponding to each confining pattern. In this paper we are going to analyze the differential-difference Sawada-Kotera family constructed in \cite{adler1} (using fractional discrete Lax operators). We will find a proliferation of tau functions as well but {\it only one} can be used in order to construct the Hirota bilinear form and compute multi-soliton solution. This tau function is a `master'-one and it is factorized in terms of the others (showing consistency of singularity patterns). Also we will study two ``modifications'' of lattice Sawada-Kotera equation. Then, we give the fully discrete general Sawada-Kotera equation obtained from its bilinear form. Finally, we shall implement the {\it express method} \cite{mase-1} to all of the confining patterns, predicting that the algebraic entropy is zero in all cases.

\section{Singularity analysis}

For a {\it differential} system, the integrability from the point of view of singularities means the absence of movable critical singularities. In order to see how one can blend the continuous and discrete situations let us consider the following example, the Volterra equation (we follow the lines presented in \cite{fane}):
$$\dot u_n=u_n(u_{n+1}-u_{n-1}),$$
which can be written as a 2-point mapping:
$${\mathbb P}^1\times {\mathbb P^1} \ni (u_n,v_n)\to (u_{n+1},v_{n+1})\in {\mathbb P}^1\times {\mathbb P}^1,$$ whose points are depending on $t$:
\begin{equation}\label{mapp}
u_{n+1}=v_{n},
\end{equation}
\begin{equation}\label{mapp1}
v_{n+1}=\frac{\dot v_n}{v_n}+u_n.
\end{equation}
\begin{remark}
    We choose ${\mathbb P}^1\times {\mathbb P}^1$ instead of ${\mathbb C}^2$ because singularity analysis includes infinities. Of course, we could have chosen ${\mathbb P}^2$ as a compactification.
\end{remark}

In order to see the singularities of $u_{n+1}, v_{n+1}$, we can  start with the formal expansion in the so-called singularity manifold:
$$u_{n}(t)=\sum_{i=0}^{\infty}a_{i}(n,t)\tau(n,t)^{i+p},$$
$$v_{n}(t)=\sum_{i=0}^{\infty}\alpha_{i}(n,t)\tau(n,t)^{i+q},$$
where $p,q$ are some numbers, $\tau(n,t)$ is the singularity manifold and $a_i(n,t)$, $\alpha_j(n,t)$ are some functions. In the Kruskal ansatz (which comes from the implicit function theorem applied in a neighborhood of $\tau(n,t)=0$), we can consider $\tau(n,t)=t-t_0(n)$ with $t_0(n)$ an arbitrary function of $n$ and accordingly, the functions $a_i, \alpha_j$ will depend {\it only} on $n$. On the other hand, since our system can be written as a 2-point mapping, the argument $n$ is no\-thing but the number of iterations.

It is obvious that if $(u_n,v_n)$ have no movable critical singularities, then the same will be true for $(u_{n+1},v_{n+1})$. Let us consider the simplest case, which is, in a neighborhood of $t$, to have a simple zero for $v_n$ and  regular $u_n$. Thus, in ${\mathbb P}^1\times {\mathbb P^1}$, the curve of coordinates
$(u_n,0)$  goes to a point with coordinates $(0,\infty)$, a situation understood as `{\it losing a degree of freedom}' (or, in the language of birational geometry, curve blow-down process). Now, because we have a mapping in $n$, the singularity confinement criterion imposes this process must be finite, and finally the initial data is recovered.
More precisely, starting as above from ($\tau=t-t_0$):
$$u_n=a_0+a_1\tau+O(\tau^2), v_n=\alpha\tau+\beta\tau^2+O(\tau^3),$$
we find from (\ref{mapp}),(\ref{mapp1}):
$$\left(
\begin{array}{c}
a_0+\cdots\\
\alpha\tau+\cdots
\end{array}\right)
\to
\left(
\begin{array}{c}
\alpha\tau+\cdots\\
\tau^{-1}+\beta/\alpha+a_0+\cdots
\end{array}\right)
\to
$$
$$
\to
\left(
\begin{array}{c}
\tau^{-1}+\beta/\alpha+a_0+\cdots\\
-\tau^{-1}+\beta/\alpha+a_0+\cdots
\end{array}\right)
\to
\left(
\begin{array}{c}
-\tau^{-1}+\beta/\alpha+a_0+\cdots\\
\gamma(a_0,\alpha,\beta,...)\tau+\cdots
\end{array}\right)
\to
\left(
\begin{array}{c}
\gamma(a_0,\alpha,\beta,...)\tau+\cdots\\
f(a_0,\alpha,\beta,...)+\cdots
\end{array}\right),
$$
where $\gamma, f$ are some finite expressions containing the parameters $a_0,\alpha, \beta,\ldots$ etc. So in a small neighbourhood of $t_0$ (where $\tau\approx 0$) we can write:
$$
\cdots\to
{\rm regular}
\to
\left(
\begin{array}{c}
a_0\\
0^1
\end{array}\right)
\to
\left(
\begin{array}{c}
0^1\\
\infty^1
\end{array}\right)
\to
\left(
\begin{array}{c}
\infty^1\\
-\infty^1
\end{array}\right)
\to
\left(
\begin{array}{c}
-\infty^1\\
0^1
\end{array}\right)
\to
\left(
\begin{array}{c}
0^1\\
f(a_0,\alpha,\beta,\ldots)
\end{array}\right)
\to {\rm regular}.
$$
So the initial curve blows down to three points and then blows up to another curve containing initial parameters (here we denote $0^p\approx \tau^p, \infty^p\approx\tau^{-p}$ for every $p>0$). In this way, the singularity confinement is satisfied. Of course, these are the {\it simplest types of singularities that we can start with}. One can start with zeros of higher order like $v\sim \alpha_0\tau^q$ and notice that in this case the length of the confined patterns will be bigger. 

The big advantage of strictly confining patterns is that they allow us to recover the Hirota bilinear form directly. Indeed, one can see immediately that for both $u_{n}, v_{n}$ the pattern is:
$$u_{n}(t): \ldots{\rm regular}\to0\to\infty\to \infty\to 0\to {\rm regular}\ldots,$$
$$v_{n-1}(t): \ldots{\rm regular}\to0\to\infty\to \infty\to 0\to {\rm regular}\ldots.$$

So we can say that exist a tau-function $F_n$ (one must not confuse tau-function specific for Hirota bilinear form with $\tau-t-t_0$) that is entire and $u_n,v_n$ are expressed as ratios of products of such functions in the form:
$$u_n=\frac{F_nF_{n-3}}{F_{n-1}F_{n-2}},$$
which is exactly the substitution that transforms Volterra equation in the Hirota bilinear form.
This one is important since the existence of general $N$-soliton solution with arbitrary parameters is characteristic of complete integrability. In this paper the existence of multisoliton solution in the bilinear form is considered the main integrability detector.

\begin{remark}
    Usually, the number of tau-functions is related to the number of 
strictly confining patterns (as in continuous case where the number of tau functions is related to the number of dominant behaviors in Painlev\'e expansion). For instance in the case of Volterra-type equation 
\begin{equation}\label{volt2}
\dot u_n=u_n(u_n-1)(u_{n+1}-u_{n-1}),  
\end{equation}
by entering through 0 and 1 as $u_{n}=0+O(t-t_0)$ or $u_n=1+O(t-t_0)$
we have two singularity patterns (``$*$'' means finite generic value)
$$*\to 0^1\to\infty^1\to 1\to *,$$
$$*\to 1\to\infty^1\to 0^1\to *,$$
which imposes two tau-functions in the relation \cite{conte}
$u_n=1-\alpha G_{n-1}F_{n+1}/G_nF_n=\beta G_{n+1}F_{n-1}/G_n F_n$ ($\alpha, \beta$ constants).
\end{remark}
\begin{remark}
 Also, another possible singular behavior appears starting from a pole of $v$, namely $v_n=\alpha/\tau+\beta+\gamma\tau+O(\tau^2)$. But in this case one obtains the so-called `anti-confining' (or weakly confining pattern), namely all forward and backward iterations contain only points as was shown in \cite{fane}. It is not clear what is the relation of an anticonfining pattern with Hirota bilinear formalism.
\end{remark}

\subsection{Singularities in higher dimensional systems}

In higher dimensions the situation is more complicated. Let us consider initially the case of pure discrete systems given by the birational mapping (we follow the description from \cite{fanee})
$$f:(\mathbb P^1)^N\to (\mathbb P^1)^N:\quad (x_1,\ldots,x_N)=(\overline{x_1},\ldots,\overline{x_N}).$$

Now, suppose there is some hypersurface $D\subset (\mathbb P^1)^N$  which is contracted by $f$ to a lower-dimensional subvariety. This contraction constitutes the kind of loss of memory of initial data, and it is called a singularity of $f$ . In the language of birational geometry, this is
nothing but the hypersurface $D$ being blown down by $f$ to the subvariety $f(D)$.
Let us introduce the set of contracted hypersurfaces
$E(f) = \{D \subset (\mathbb P^1)^N| \det(\partial_x f)|_{D} = 0\}$,
where the vanishing of the Jacobian indicates the contraction to a lower dimensional subvariety.
The singularity constituted by contraction of $D\in E(f)$ is said to be confined if there exists an
integer $n\geq 2$ such that $f^n(D)$ is of codimension 1. Then the coordinates of $D$ are recovered
in those of another hypersurface $f^n(D)$, and in this sense the memory of initial conditions is
recovered. If the singularity corresponding to the contraction of $D$ is confined for {\it every} $D$ in $E(f)$ we say that $f$ satisfies the singularity confinement criterion (note that the existence of a possible confined singularity of $f : (\mathbb P^1)^N\to (\mathbb P^1)^N$ implies, that $f$ is not algebraically stable)

Now let us consider the case of a general differential-difference system (with coordinates depending on the continuous variable $t$) written as a mapping of the projective space:
$$({\mathbb P}^1)^N \ni (x_1,\ldots,x_N)\to (\overline{x_1},\ldots,\overline{x_N})\in ({\mathbb P}^1)^N, \quad x_i\equiv x_i(n,t),\quad {\overline x_i}\equiv x_i(n+1,t), \forall i=1,\ldots
,N$$
and given by the following birational form $({\mathbf x_{n}(t)}\equiv (x_1(n,t),x_2(n,t),\ldots,x_N(n,t))^T),$
$${\mathbf x_{n+1}(t)}={\mathbf F}({\mathbf x_n(t)},\partial_t{\mathbf x_n(t)}).$$

%$$\overline{x_N}=F_N(x_1,...,x_N;\partial_t x_1,...,\partial_t x_N)$$
%The dynamics is given by the composition of the mapping $\phi$. In this way the correspondence $x_i(t)\to \overline {x_i(t)}$ is the same as $x_i(n,t)\to x_i(n+1,t)$

Consider that in the `time' variable $t$ the coordinates can be expressed in general by some Laurent series of the form:
$$x_i(n,t)=\sum_{k\geq 0}\alpha_{ik}(n)(t-t_0)^{k+p_i}.$$
Let $i\in{1,2,...,N}$ such that $x_i(n,t)=a_i+O(t-t_0)$ and $x_j(n,t)=A_j(n)+O(t-t_0)$ for any $j\neq i$ (i.e. $a_i$ is a fixed entry).

Now suppose that the hypersurface $D=(A_1,A_2,\ldots,A_{i-1},a_i,A_{i+1},\ldots,A_N)$ is contracted by ${\mathbf F}$ to a lower-dimensional
subvariety. This contraction constitutes a loss of memory of initial data, and represents a {\it singularity} of ${\mathbf F}$.
Several problems appear now. The first one is that we cannot define the exceptional set $E({\mathbf F}
)$ because we have derivatives. Accordingly, we cannot say anything about algebraic stability. Secondly, lower dimensionality (or losing a degree of freedom) is realized either by the appearance of some fixed entries in the coordinates or by some relations among coordinates. In the differential-difference context it is extremely hard to control the relations among coordinates because not only the dominant terms of Laurent series appear, but also the dominant terms of the derivatives. As iterations go further, more and more coefficients of the series are involved. That is why it is not clear how to define rigorously the singularity confinement here. However, since we are interested mainly in the positions of tau functions and the application of the `express method' we consider {\it only the presence of fixed entries in the subvarieties as defining `singularities'}. Accordingly we are going to focus only on the {\it to fixed entries confined patterns} and all other patterns (anti-confining, cyclic confining) will be discarded. `Regular' entries are those who do not contain fixed entries.

\section{Sawada-Kotera type lattice equations}
The equations under consideration are the following:

Ordinary differential-difference Sawada-Kotera \cite{hir} (we denote it SK1):
$$
\partial_tv_{n}=v_n^2(v_{n+2}v_{n+1}-v_{n-1}v_{n-2})-v_n(v_{n+1}-v_{n-1}).
$$
We will study it together with one modification of it (SK2) (the simplest one in the list of \cite{adler1}):

$$
\partial_tu_{n}=u_{n+1}u_n^3u_{n-1}(u_{n+2}u_{n+1}-u_{n-1}u_{n-2})-u_n^2(u_{n+1}-u_{n-1}).
$$
Then we will study the general case (order $2m$) and call it (SKg):

$$\partial_t v_{n}=v_n^2(v_{n+m}v_{n+m-1}\cdots v_{n+1}-v_{n-1}v_{n-2}\cdots v_{n-m})-v_n(v_{n+m-1}\cdots v_{n+1}-v_{n-1}v_{n-2}\cdots v_{n-m+1}).$$
In the final part we will discuss a more complicated modification based on M\"obius invariance (SK3):
$$\partial_t x_{n}=(x_n+1)\left(\frac{x_{n+2}x_n(x_{n+1}+1)^2}{x_{n+1}}-\frac{x_{n-2}x_n(x_{n-1}+1)^2}{x_{n-1}}+(2x_n+1)(x_{n+1}-x_{n-1})\right).$$
We mention that all these equations have classical Sawada-Kotera as continuum limit:
$$U_{t} = U_{xxxxx} + 5UU_{xxx} + 5U_xU_{xx} + 5U^2U_x.$$

\subsection{Patterns and tau functions}
\subsubsection{SK1 equation}

Let us write the SK1 equation as a dynamical system: 

$$\phi:({\mathbb P}^1)^4\to ({\mathbb P}^1)^4,\quad (v_1,v_2,v_3,v_4)\to (\bar{v_1},\bar {v_2},\bar{v_3}, \bar{v_4}),$$

$$\bar{v_1}=v_2,$$
$$\bar{v_2}=v_3,$$
$$\bar{v_3}=v_4,$$
$$\bar{v_4}=\frac{-v_3v_2+v_3^2v_2v_1+v_3v_4+\partial_t v_{3}}{v_3^2v_4}.$$
    
%We also have the inverse mapping:
%$$\phi^{-1}:({\mathbb P}^1)^4\to ({\mathbb P}^1)^4,\quad (v_1,v_2,v_3,v_4)\to (\underline{v_1},\underline{v_2},\underline{v_3}, \underline{v_4}),$$

%$$\underline{v_1}=\frac{v_2v_1-v_2v_3+v_2^2v_3v_4-v_{2,t}}{v_2^2v_1},$$
%$$\underline{v_2}=v_1,$$
%$$\underline{v_3}=v_2,$$
%$$\underline{v_4}=v_3.$$

One can identify the two possible entrances which may produce singularities (with fixed entries $0$ , $\infty$, etc) in the direct mapping $\phi$ $(
v_3=0,v_4=0)$.
We will analyse all of them:\\
For $v_3 =0$ we will find for the forward evolution (given by iteration of $\phi$) the following pattern:
$$
\left(
\begin{array}{c}
a_1\\
a_2\\
0^1\\
a_4
\end{array}\right)
\to
\left(
\begin{array}{c}
*\\
0^1\\
*\\
\infty^2
\end{array}\right)\to
\left(
\begin{array}{c}
0^1\\
*\\
\infty^2\\
*
\end{array}\right)\to
\left(
\begin{array}{c}
*\\
\infty^2\\
*\\
0^1
\end{array}\right)\to
\left(
\begin{array}{c}
\infty^2\\
*\\
0^1\\
*
\end{array}\right)\to
$$
$$
\left(
\begin{array}{c}
*\\
0^1\\
*\\
*
\end{array}\right)\to
\left(
\begin{array}{c}
0^1\\
*\\
*\\
*
\end{array}\right).
$$

%For the backward evolution (iteration of $\phi^{-1}$)
%$$
%{\rm regular} \to
%\left(
%\begin{array}{c}
%*\\
%*\\
%*\\
%0
%\end{array}\right)
%\to\left(
%\begin{array}{c}
%a_1\\
%a_2\\
%0\\
%a_4
%\end{array}\right).
%$$

The next pattern given by $(v_4=0)$ is the following:
$$
\left(
\begin{array}{c}
 a_1 \\
 a_2 \\
 a_3 \\
 0^1  \\
\end{array}
\right)\to
\left(
\begin{array}{c}
 * \\
 * \\
 0^1 \\
 \infty^1
\end{array}
\right)\to
\left(
\begin{array}{c}
 * \\
 0^1 \\
 \infty^1  \\
 \infty^1  \\
\end{array}
\right)\to
\left(
\begin{array}{c}
 0^1 \\
 \infty^1  \\
 \infty^1  \\
 0^1 \\
\end{array}
\right)\to
\left(
\begin{array}{c}
 \infty^1  \\
 \infty^1  \\
 0^1 \\
 *
\end{array}
\right)\to
\left(
\begin{array}{c}
 \infty^1  \\
 0^1 \\
 *  \\
 *  \\
\end{array}
\right)\to
\left(
\begin{array}{c}
 0^1 \\
 * \\
 *  \\
 * \\
\end{array}
\right).
$$

%For the backward evolution (iteration of $\phi^{-1}$)
%$$
%{\rm regular} \to
%\left(
%\begin{array}{c}
%*\\
%*\\
%*\\
%*
%\end{array}\right)
%\to\left(
%\begin{array}{c}
%a_1\\
%a_2\\
%a_3\\
%0
%\end{array}\right).
%$$

So we have two confined singularity patterns with fixed entries which must be compatible. The one starting with $v_3=0^1$:
\begin{equation}
0^1\to *\to \infty^2\to *\to 0^1,
\end{equation}
as well as the one starting with $v_4=0^1$:
\begin{equation}\label{sk12}
0^1\to\infty^1\to\infty^1\to 0^1.
\end{equation}
However since $v_4(n,t)=v_3(n+1,t)$, the dependent variable (and its shifted value) produces two singularity pattern by entering through the {\it same value, } $0^1$. This is in contrast with the situation discussed for equation (\ref{volt2}) where the dependent variable produces two singularity patterns by entering through {\it two different} values.
Accordingly the first singularity pattern gives the Hirota bilinear substitution
$$v_n=\frac{f_{n-2}f_{n+2}}{f_n^2},$$
while the second gives
$$v_n=\frac{F_{n-1}F_{n+2}}{F_nF_{n+1}},$$
which means that we have {\it two possible tau-functions $f_n$ and $F_n$}. However, one can see immediately that $F_n=f_nf_{n+1}$ and the two patterns are indeed compatible. We use the second pattern tau function to construct the Hirota bilinear form. Defining the Hirota operator $D_ta\cdot b\equiv a_tb-ab_t$, we obtain:
$$\frac{(D_t F_{n-1}\cdot F_n)F_{n+1}F_{n+2}-(D_t F_{n+1}\cdot F_{n+2})F_{n-1}F_n}{F_n^2F_{n+1}^2}=\frac{F_{n-1}^2F_{n+4}}{F_nF_{n+1}^2}-\frac{F_{n+2}^2F_{n-3}}{F_n^2F_{n+1}}-\frac{F_{n+3}F_{n-1}}{F_{n+1}^2}+\frac{F_{n+2}F_{n-2}}{F_n^2}$$
which turns into:
$$(D_t F_{n-1}\cdot F_n+F_{n-3}F_{n+2}-F_{n+1}F_{n-2})F_{n+1}F_{n+2}=(D_t F_{n+1}\cdot F_{n+2}+F_{n-1}F_{n+4}-F_{n+2}F_n)F_{n-1}F_n,$$
Again one can see that the first factor in the rhs is the double up-shift of the first factor in the lhs, so we can write, accordingly:
$$D_t F_{n-1}\cdot F_n+F_{n-3}F_{n+2}-F_{n+1}F_{n-2}=\beta F_nF_{n-1}.$$
The presence of integration constant shows we have many possible solutions (the constant being related to the initial/boundary conditions). But we are interested only in the multi-soliton solution (for integrability reasons) and, being a particular solution, it requires a particular value for the constant $\beta$.

\noindent Indeed, for $\beta=0$ we find the following multi-soliton solution
\begin{equation}\label{sol}
 F_n(t)=\sum_{\mu_1,\ldots ,\mu_M\in\{0,1\}}\exp\left(\sum_{i=1}^M\mu_i(k_in+\omega_it)+\sum_{i<j}^MA_{ij}\mu_i\mu_j\right),
\end{equation}
with the dispersion relation and interaction phase given by:
$$
\omega_i=2\sinh(2k_i),
$$
$$
\exp A_{ij}=\frac{(e^{k_i}-e^{k_j})^2(e^{k_i}+e^{k_j})}{(e^{k_i+k_j}-1)^2(e^{k_i+k_j}-1)}.
$$

\begin{remark}
    In the case of Bogoyavlenski lattice of the form
$$\partial_tv_{n}=v_n^2(v_{n+2}v_{n+1}-v_{n-1}v_{n-2}),$$
we have two patterns as well. The second one is the same as (\ref{sk12}), but the first one is very asymmetric \cite{fane} i.e.
$$0^1\to *\to \infty^2\to 0^1\to \infty^2\to 0^2,$$
and compatibility imposes the following more complicated relation between tau-functions $F_n=f_{n-1}f_nf_{n+2}^2$.

\end{remark}

\subsubsection{SK2 equation}

For SK2 equation we write it as:

$$\phi:({\mathbb P}^1)^4\to ({\mathbb P}^1)^4,\quad (u_1,u_2,u_3,u_4)\to (\bar{u_1},\bar {u_2},\bar{u_3}, \bar{u_4}),$$

$$\bar{u_1}=u_2,$$
$$\bar{u_2}=u_3,$$
$$\bar{u_3}=u_4,$$
$$\bar{u_4}=\frac{-u_3^2u_2+u_3^2u_4+u_3^3u_2^2u_1u_4+\partial_tu_{3}}{u_2u_3^3u_4^2}.$$

One can identify the three possible entries which may produce singularities in the mapping $\phi$ $(u_2=0, u_3=0, u_4=0)$.

For $u_2 =0$ we obtain an anti-confining pattern. The same with $u_3=0$

The pattern for ($u_4=0$) is confining:
$$
\left(
\begin{array}{c}
a_1\\
a_2\\
a_3\\
0^1
\end{array}\right)
\to
\left(
\begin{array}{c}
*\\
*\\
0^1\\
\infty^2
\end{array}\right)
\to
\left(
\begin{array}{c}
*\\
0^1\\
\infty^2\\
0^1
\end{array}\right)
\to
\left(
\begin{array}{c}
0^1\\
\infty^2\\
0^1\\
*
\end{array}\right)
\to
\left(
\begin{array}{c}
\infty^2\\
0^1\\
*\\
*
\end{array}\right)
\to
\left(
\begin{array}{c}
0^1\\
*\\
*\\
*
\end{array}\right).
%\to
%\left(
%\begin{array}{c}
%*\\
%*\\
%*\\
%*
%\end{array}\right)
$$
From it we get:
$$u_n=F_{n-1}F_{n+1}/F_{n}^2.$$
%Let us make the following notation: $F_n=F, F_{n+k}=F_k, F_{n-k}=F_{-k}, \forall k$ and we are going to use also the Hirota bilinear operator $D_t^p a\cdot b=(\partial_t-\partial_{t'})^pa(t)b(t')|_{t=t'}.$

Introducing in the equation we obtain:
$$\frac{(D_t F_{n+1}\cdot F_n)F_{n-1}-(D_t F_n\cdot F_{n-1})F_{n+1}}{F_n^3}=\frac{F_{n-1}F_{n-2}F_{n+3}}{F_n^3}-\frac{F_{n+2}F_{n+1}F_{n-3}}{F_n^3}-\frac{F_{n+2}F_{n-1}^2}{F_n^3}+\frac{F_{n+1}^2F_{n-2}}{F_n^3},$$
which goes to:
$$(D_t F_{n+1}\cdot F_n-F_{n-2}F_{n+3}+F_{n+2}F_{n-1})F_{n-1}=(D_t F_n\cdot F_{n-1}-F_{n+2}F_{n-3}+F_{n-2}F_{n+1})F_{n+1}.$$
One can immediately see that the first factor in the lhs is the up-shift of the first factor in the rhs. Accordingly we can consider:
\begin{equation}
D_t F_{n+1}\cdot F_n-F_{n-2}F_{n+3}+F_{n+2}F_{n-1}=\alpha F_{n+1} F_n,
\end{equation}
where $\alpha=0$ yields the multi-soliton solution (\ref{sol}).

\section{General case, (SKg) equation}
%\subsection{Singularity analysis}
In this section we will try to apply singularity confinement to the case of general higher order lattice Sawada-Kotera of order $2m$ constructed by Adler \cite{adler1} (using fractional Lax operators) namely ($m=2$ is ordinary lattice Sawada-Kotera)
$$\partial_t v_{n}=v_n^2(v_{n+m}v_{n+m-1}\cdots v_{n+1}-v_{n-1}v_{n-2}\cdots v_{n-m})-v_n(v_{n+m-1}\cdots v_{n+1}-v_{n-1}v_{n-2}\cdots v_{n-m+1})$$
It was shown that this general equation has as its continuous limit the Sawada-Kotera equation. The integrability was shown, based on the compatibility of the following Lax pair:
$$v_n\psi_{n+m+1}-\psi_{n+m}+\lambda(\psi_{n+1}-v_n\psi_n)=0,$$
$$\partial_t\psi_n-v_{n-1}\cdots v_{n-m}(\lambda\psi_{n-m}-\lambda^{-1}\psi_{n+m})=0.$$

Let us consider the case $m=3$:
$$\bar{v_1}=v_2,$$
$$\bar{v_2}=v_3,$$
$$\bar{v_3}=v_4,$$
$$\bar{v_4}=v_5,$$
$$\bar{v_5}=v_6,$$
$$\bar{v_6}=\frac{-v_4v_3v_2+v_4^2v_3v_2v_1+\partial_t v_{4}+v_4v_5v_6}{v_4^2v_5v_6}.$$
Here we have three possible sources of singularities ($v_4, v_5, v_6=0$). All of them give strictly confining patterns:

For $v_4=0$ we get:
$$
\left(
\begin{array}{c}
a_1\\
a_2\\
a_3\\
0^1\\
a_5\\
a_6
\end{array}\right)
\to
\left(
\begin{array}{c}
*\\
*\\
0^1\\
*\\
*\\
\infty^2
\end{array}\right)
\to
\left(
\begin{array}{c}
*\\
0\\
*\\
*\\
\infty^2\\
*
\end{array}\right)
\to
\left(
\begin{array}{c}
0^1\\
*\\
*\\
\infty^2\\
*\\
*
\end{array}\right)
\to
\left(
\begin{array}{c}
*\\
*\\
\infty^2\\
*\\
*\\
0^1
\end{array}\right)
\to
$$

$$
\left(
\begin{array}{c}
*\\
\infty^2\\
*\\
*\\
0^1\\
*
\end{array}\right)
\to
\left(
\begin{array}{c}
\infty^2\\
*\\
*\\
0^1\\
*\\
*
\end{array}\right)
\to
\left(
\begin{array}{c}
*\\
*\\
0^1\\
*\\
*\\
*
\end{array}\right)
\to
\left(
\begin{array}{c}
*\\
0^1\\
*\\
*\\
*\\
*
\end{array}\right).
$$

The next singularity may enter through $v_5=0$ and produces the following confining pattern:

$$
\left(
\begin{array}{c}
a_1\\
a_2\\
a_3\\
a_4\\
0^1\\
a_6
\end{array}\right)
\to
\left(
\begin{array}{c}
*\\
*\\
*\\
0^1\\
*\\
\infty^1
\end{array}\right)
\to
\left(
\begin{array}{c}
*\\
*\\
0^1\\
*\\
\infty^1\\
\infty^1
\end{array}\right)
\to
\left(
\begin{array}{c}
*\\
0^1\\
*\\
\infty^1\\
\infty^1\\
*
\end{array}\right)
\to
\left(
\begin{array}{c}
0^1\\
*\\
\infty^1\\
\infty^1\\
*\\
0^1
\end{array}\right)
\to
$$

$$
\to
\left(
\begin{array}{c}
*\\
\infty^1\\
\infty^1\\
*\\
0^1\\
*
\end{array}\right)
\to
\left(
\begin{array}{c}
\infty^1\\
\infty^1\\
*\\
0^1\\
*\\
*
\end{array}\right)
\to
\left(
\begin{array}{c}
\infty^1\\
*\\
0^1\\
*\\
*\\
*
\end{array}\right)
\to
\left(
\begin{array}{c}
*\\
0^1\\
*\\
*\\
*\\
*
\end{array}\right)
\to
\left(
\begin{array}{c}
0^1\\
*\\
*\\
*\\
*\\
*
\end{array}\right).
$$

The last possibility is to enter through $v_6=0$. Here we have again a strictly confining pattern:

$$
\left(
\begin{array}{c}
a_1\\
a_2\\
a_3\\
a_4\\
a_5\\
0^1
\end{array}\right)
\to
\left(
\begin{array}{c}
*\\
*\\
*\\
*\\
0^1\\
\infty^1
\end{array}\right)
\to
\left(
\begin{array}{c}
*\\
*\\
*\\
0^1\\
\infty^1\\
*
\end{array}\right)
\to
\left(
\begin{array}{c}
*\\
*\\
0^1\\
\infty^1\\
*\\
\infty^1
\end{array}\right)
\to
\left(
\begin{array}{c}
*\\
0^1\\
\infty^1\\
*\\
\infty^1\\
0^1
\end{array}\right)
\to
$$
$$
\to
\left(
\begin{array}{c}
0^1\\
\infty^1\\
*\\
\infty^1\\
0^1\\
*
\end{array}\right)
\to
\left(
\begin{array}{c}
\infty^1\\
*\\
\infty^1\\
0^1\\
*\\
*
\end{array}\right)
\to
\left(
\begin{array}{c}
*\\
\infty^1\\
0^1\\
*\\
*\\
*
\end{array}\right)
\to
\left(
\begin{array}{c}
\infty^1\\
0^1\\
*\\
*\\
*\\
*
\end{array}\right)
\to
\left(
\begin{array}{c}
0^1\\
*\\
*\\
*\\
*\\
*
\end{array}\right).
$$

Thus we have three confining patterns. For $v_4=0$ we have the following orbit

$$0^1\to *\to *\to\infty^2\to *\to *\to 0^1.$$

For $v_5=0$:

$$0^1\to *\to \infty^1\to\infty^1\to *\to 0^1,$$

and for $v_6=0$:
$$0^1\to \infty^1\to *\to \infty^1\to 0^1,$$
giving the following {\it proliferation of tau-functions}:

$$v_n=\frac{F_{n-3}F_{n+3}}{F_n^2},\quad v_n=\frac{f_{n-2}f_{n+3}}{f_nf_{n+1}}, \quad v_n=\frac{\phi_{n-1}\phi_{n+3}}{\phi_n\phi_{n+2}}.$$

The compatibility can be seen in:
$$\phi_n=f_nf_{n-1}, \quad \phi_n=F_nF_{n-1}F_{n-2}.$$
The bilinear form can be done immediately with the substitution:
$$v_n=\frac{\phi_{n+1}\phi_{n+3}}{\phi_n\phi_{n+2}},$$
and it is:
$$D_t\phi_{n+1}\cdot \phi_n=\phi_{n+3}\phi_{n-2}-\phi_{n-3}\phi_{n+4}.$$

\begin{remark}
    This is exactly the bilinear form found in \cite{adler1}. All the other substitutions give multi-linear equations.
\end{remark}

This situation can be easily generalized to any $m$; namely, we will have the following $m$ patterns corresponding to $m$-factors at the denominator (one of it is $v_n^2$ which will give the first pattern containing $\infty^2$; this $\infty^2$ will spread in a pair of $\infty^1$ which, in the next patterns `move' towards both extremities until the extremal zeros). More precisely:
\begin{equation}
0^1\to *\underset{(m-1)-{\rm times}}{\cdots }\to *\to\infty^2\to *\underset{(m-1)-{\rm times}}{\cdots }\to *\to0^1,
\end{equation}
\begin{equation}
0^1\to *\underset{(m-2)-{\rm times}}{\cdots }\to *\to\infty^1\to\infty^1\to *\underset{(m-2)-{\rm times}}{\cdots }\to *\to 0^1,
\end{equation}
\vskip 0.2cm
$$\cdots \cdots \cdots \cdots \cdots \cdots $$
\vskip 0.2cm
\begin{equation}
0^1\to\infty^1\to *\underset{(m-2)-{\rm times}}{\cdots }\to *\to\infty^1\to 0^1.
\end{equation}
This last pattern will give the Hirota bilinear substitution together with the Hirota bilinear form:
\begin{equation}\label{adler-bil}
v_n=\frac{\phi_{n+m+1}\phi_n}{\phi_{n+m}\phi_{n+1}},\quad D_t\phi_{n+1}\cdot \phi_n=\phi_{n+m}\phi_{n-m+1}-\phi_{n-m}\phi_{n+m+1}.
\end{equation}

The other tau-functions have essentially the same relation with $\phi$, namely (we change notation and use the discrete index $n$; we denote the tau-functions from bottom to top as $f_1, f_2, \ldots , f_m$):

$$\phi_n=f_{1,n}f_{1,n-1}=f_{2,n}f_{2,n-1}f_{2,n-2}=\cdots =f_{m,n}f_{m,n-1}\ldots f_{m,n-m+1}.$$

%***********************************************
%\vskip 1cm
%The first term of SK1 can be generalised to the following completely integrable form:
%$$\dot u_n=u_{n}^{N+1}\prod_{i=1}^{N-1}(u_{n+i}u_{n-i})^{N-i}(\prod_{j=1}^Nu_{n+j}-\prod_{j=1}^Nu_{n-j})$$

%The bilinear substitution is:
%$$u_n=F_{n-1}F_{n+1}/F_n^2$$

%With this bilinear substitution the Hirota bilinear form is:
%$$D_t F\cdot F_{1}+F_{N+1}F_{-N}-FF_{1}=0$$
%This is again a Hirota-Miwa type equation which has N-soliton solution.

%The general form of the first term of SK2 is the well-known Bogoyavlensky lattice:
%$$\dot u_n=u_{n}^{2}(\prod_{j=1}^Nu_{n+j}-\prod_{j=1}^Nu_{n-j})$$
%with
%\begin{equation}\nonumber
% u_n=\frac{F_{-1}F_{N}}{FF_{N-1}}.
%\end{equation}
%which gives the following relations:
%$$u^2\prod_{i=1}^Nu_{i}=\frac{F_{-1}^2F_{2N}F}{F^2F_{N-1}^2}$$
%$$u^2\prod_{i=1}^Nu_{-i}=\frac{F_{N}^2F_{-N-1}F_{N-1}}{F^2F_{N-1}^2}$$
%The bilinear form will be
%\begin{equation}\label{bmB2}
%D_t F_{-1}\cdot F-F_{N}F_{-N-1}-FF_{-1}=0
%\end{equation}
%**************************************************
\subsection{Time discretization of SKg: bilinear approach}
We shall use the bilinear form to also discretize in time the above equations. It is easy to discretize the bilinear form. The main problem appears when one has to recover the nonlinear form.

Let us make some notations. When we discretize in time and space, $v(t,n)\to v(\nu,n)\equiv v_{\nu,n}$. We thus make the following notations:
$$F_{\nu n}=F, F_{\nu+1,n}=\tilde {F_n}, F_{\nu+2,n}=\tilde{\tilde{F_n}}, {\rm etc}.$$
The Hirota bilinear operator will be discretized in a standard way by replacing derivatives with finite differences ($\delta$ is the discretization step):
$$D_ta\cdot b\equiv a_tb-ab_t\to\frac{1}{\delta}((a(t+\delta)-a(t))b(t)-a(t)(b(t+\delta)-b(t)))= \frac{1}{\delta}(a(t+\delta)b(t)-a(t)b(t+\delta)).$$
If we replace $t$ by $\nu\delta$ we get:
$$D_ta\cdot b\to\frac{1}{\delta}(a(\nu+1)b(\nu)-a(\nu)b(\nu+1))\equiv \frac{1}{\delta}(\tilde{a}b-a\tilde{b}).$$

Let us take the bilinear form (\ref{adler-bil}):
$$D_t F_{n+1}\cdot F_{n}-F_{n+m}F_{n-m+1}+F_{n-m}F_{n+m+1}=0,$$
and consider its time-discretization namely $F=F_n$. Replacing Hirota bilinear operator we find 
\begin{equation}
\tilde{F}_{n+1}F_n-F_{n+1}\tilde{F_{n}}-\delta \tilde{F}_{n+m}F_{n-m+1}+\delta F_{n-m}\tilde{F}_{n+m+1}=0.
\end{equation}
We shifted one variable of each term with a tilde because any Hirota bilinear equation must be {\it gauge-invariant}, namely $F(\nu,n)\to F(\nu,n) e^{an+b\nu}$ for any constants $a,b$.

\begin{remark}
    The discretized bilinear form is {\it not} automatically integrable. One has to check the existence of multisoliton solution (since it is extremely restrictive the existence of 3-soliton solution is usually enough \cite{jarmo-bil}). Indeed we have:
$$F(\nu,n)=\sum_{\mu_1,\ldots,\mu_N\in\{0,1\}}\left(\prod_{i=1}^Np_i^{\mu_i n}q_{i}^{\mu_i\nu}\prod_{i<j}^NA_{ij}^{\mu_i\mu_j}\right),$$
where:
$$q_i=\frac{p_i^{-m}(\delta+p_i^m)}{1+\delta p_i^m},\quad A_{ij}=\frac{(p_i-p_j)(p_i^m-p_j^m)}{(p_ip_j-1)(p_i^mp_j^m-1)}.$$
\end{remark}

Considering the substitution for (\ref{adler-bil}) we discretize in the form:
\begin{equation}\label{sub}
v=\frac{\tilde{F}_{n+m+1}F_{n}}{\tilde{F}_{n+m}F_{n+1}}.
\end{equation}

In order to find the nonlinear form we divide the bilinear equation by $\tilde{F}_{n}F_{n+1}$ and obtain:
\begin{equation}
\frac{\tilde{F}_{n+1}F_n}{\tilde{F}_{n}F_{n+1}}-1-\delta\frac{\tilde{F}_{n+m}F_{n-m+1}}{\tilde{F}_{n}F_{n+1}}+\delta\frac{F_{n-m}\tilde{F}_{n+m+1}}{\tilde{F}_{n}F_{n+1}}=0.
\end{equation}
We have to express these three terms as combinations of various shifts of (\ref{sub}).
Let us denote: 
$$K_n=\frac{\tilde{F}_{n+1}F_n}{\tilde{F}_{n}F_{n+1}}\cdot$$
We will drop the subscript of $K_n$ in the following equations. One can see that:
$$\frac{\tilde{F}_{n+m}F_{n-m+1}}{\tilde{F}_{n}F_{n+1}}=K\prod_{i=1}^{m-1} v_{n-i},$$
$$\frac{F_{n-m}\tilde{F}_{n+m+1}}{\tilde{F}_{n}F_{n+1}}=\prod_{i=0}^m v_{n-i}.$$

Accordingly:
$$K-1-\delta K \prod_{i=1}^{m-1} v_{n-i}+\delta \prod_{i=0}^m v_{n-i}=0,$$ which gives
\begin{equation}
K=\frac{1-\delta \prod_{i=0}^m v_{n-i}}{1-\delta \prod_{i=1}^{m-1} v_{n-i}}\cdot
\end{equation}
On the other hand we have the relation:
$$\frac{\tilde{v}}{v}=\frac{\tilde{K}_{n+m}}{K}\cdot$$
Therefore, up-shifting and down-shifting $K$ from (13) we obtain the nonlinear form written explicitly with all indices:
$$\frac{v_{\nu+1,n}}{v_{\nu,n}}=\frac{(1-\delta \prod_{i=0}^{m} v_{\nu+1,n+m-i})(1-\delta \prod_{i=1}^{m-1} v_{\nu,n-i})}{(1-\delta \prod_{i=1}^{m-1} v_{\nu+1,n+m-i})(1-\delta \prod_{i=0}^{m} v_{\nu,n+m-i})}\cdot$$

%For the case of SK1 it is more difficult. The bilinear substitution will be
%$$u_{0,0}=\frac{F_{0,-1}F_{1,1}}{F_{0,0}F_{1,0}}$$
%The bilinear form is the same.  Also 
%$$\frac{K_{1,0}}{K_{0,-1}}=\frac{u_{1,0}}{u_{0,0}}$$
%We use the same machinery but we cannot solve for $K_{0,0}$. So we are forced to write
%$$K_{0,0}-1+\delta u_{0,0}u_{0,-1}=\delta \frac{F_{0,-2}F_{1,3}}{F_{0,1}F_{1,0}}$$
%The product in the rhs cannot be written (or I am too stupid to see) in terms of shifts of $u_{0,0}$. But the ratio of un-shift in $m$ and down-shift in $n$, can. So, the final nonlinear form is a system in $K$ and $u$

%$$\frac{K_{1,0}-1+\delta u_{1,0}u_{1,1}}{K_{0,-1}-1+\delta u_{0,0}u_{0,-1}}=\frac{ u_{1,0}u_{1,1}u_{1,2}}{u_{0,0}u_{0,-1}u_{0,-2}}$$
%$$\frac{K_{1,0}}{K_{0,-1}}=\frac{u_{1,0}}{u_{0,0}}$$

\section{Express method}
Veselov \cite{veselov} realized that integrability in discrete settings has an essential correlation with the weak growth of certain characteristics, based on a statement by Arnold \cite{arnold}, who introduced
the notion of complexity for mappings on the plane. The latter is defined as the number of intersection
points of a fixed curve with the images of a second curve under the $n$-th iteration of the mapping. 
Bellon and Viallet \cite{belon} made this idea more precise by considering the
limit of the degree of iterates of the mapping when $n \to \infty$, introducing the quantity
$S = \lim_{n\to\infty}(\log d_n)/n$,
which is called algebraic entropy ($\lambda = \exp(S)$ is often referred to as the dynamical degree of the mapping). A strictly positive value for $S$ (corresponding to a dynamical degree greater than $1$) is an indication of non-integrability, while integrability means zero algebraic entropy (and
dynamical degree equal to 1).
Singularity patterns {\it can} provide the complexity growth. In the case of two dimensional discrete mappings it was shown by Halburd in \cite{halburd} that, from the structure of confining patterns, one can estimate it together with the value of the algebraic entropy as well. However one has to take into account all open patterns (strictly confining) and cyclic patterns. Later, a simplified version called `express method' was found \cite{mase-1} which provides an algorithm for computing only the algebraic entropy.
The algorithm is the following: For a given singularity pattern one associates a monomial $c_j \lambda^{j-1}$
with each entry in the pattern, where $j$ is the position of each entry and $c_j$ is $(\pm 1)\times$(exponent of the
$j$-entry), depending if it is finite (plus sign) or infinite (minus sign). The logarithm of the largest root of sum of these monomials gives the algebraic entropy. One can see that the polynomial (formed by these monomials) is relying on the fixed entries of the singularities. The Halburd's method was extended to higher dimensional mappings as well in \cite{fane-ralph}.

Of course one can wonder if the simplified `express method'  can be extended to higher-dimensional differential-difference case. We do not have a clear cut answer but inasmuch as our patterns are relying only on fixed entries we can try to use it.

%\subsection{SKg equation}

Let us consider patterns in (Skg):
$$
0^1\to *\underset{(m-1)-{\rm times}}{\cdots}\to *\to\infty^2\to *\underset{(m-1)-{\rm times}}{\cdots}\to *\to0^1,
$$

$$
0^1\to *\underset{(m-2)-{\rm times}}{\cdots}\to *\to\infty^1\to\infty^1\to *\underset{(m-2)-{\rm times}}{\cdots}\to *\to 0^1,
$$
$$
0^1\to\infty^1\to *\underset{(m-2)-{\rm times}}{\cdots}\to *\to\infty^1\to 0^1.
$$

The polynomial associated with the first pattern is $1-2\lambda^m+\lambda^{2m}$. Its maximum root is $\lambda=1$ - compatible with integrability.\\
The second pattern leads to the polynomial $1-\lambda^{m-1}-\lambda^m+\lambda^{2m-1}=(1-\lambda^m)(1-\lambda^{m-1})$. The modulus of all roots is 1, again compatible with integrability.\\
The last polynomial is $1-\lambda-\lambda^m+\lambda^{m+1}=(1-\lambda)(1-\lambda^m)$, which again has the modulus of all roots equal to 1, which is compatible with integrability.\\
An intermediate confining pattern in the sequence writes as:
$$
0^1\to *\underset{k-{\rm times}}{\cdots}\to *\to\infty^1\to*\underset{(m-2-k)-{\rm times}}{\cdots}\to *\to\infty^1\to *\underset{k-{\rm times}}{\cdots}\to *\to 0^1.
$$
Its associated polynomial is $1-\lambda^{k+1}-\lambda^{m}+\lambda^{m+1+k}=(1-\lambda^m)(1-\lambda^{k+1})$. The same conclusion as before also holds here.

\subsection{SK3 equation}
In this case, the situation is more complicated. First of all, this mapping is a modification of the Sawada-Kotera equation that is invariant to Mobius transformations. The resulting SK3 equation is related to a {\it schwarzian}-type of Bogoyavlenski lattice which is also related to Sawada-Kotera equation. We expect a more complicated singularity structure.

Indeed, let us write the equation as a dynamical system:
$$\phi:({\mathbb P}^1)^4\to ({\mathbb P}^1)^4,\quad (x_1,x_2,x_3,x_4)\to (\bar{x_1},\bar {x_2},\bar{x_3}, \bar{x_4}),$$

$$\bar{x_1}=x_2,$$
$$\bar{x_2}=x_3,$$
$$\bar{x_3}=x_4,$$
$$\bar{x_4}=-\frac{x_4 (-x_3 x_1 (1 + x_3) (1 + x_2)^2 - x_2 \partial_t x_{3} +
    x_2 (1 + x_3) (1 + 2x_3) (x_4 - x_2))}{x_3 (1 + x_3) x_2 (1 + x_4)^2}.$$

One can identify the five possible entrances that may produce singularities in the direct mapping $\phi$ $(
x_2=0,x_3=0,x_3=-1,x_4=0,x_4=-1)$.  Cases $x_2=0, x_3=0$ are producing nonconfined patterns.

Next, we consider that the singularity enters through $x_3=-1$. We obtain the following confining pattern:

$$
\left(
\begin{array}{c}
a_1\\
a_2\\
-1\\
a_4
\end{array}\right)
\to
\left(
\begin{array}{c}
*\\
-1\\
*\\
\infty^1
\end{array}\right)
\to
\left(
\begin{array}{c}
-1\\
*\\
\infty^1\\
*
\end{array}\right)
\to
\left(
\begin{array}{c}
*\\
\infty^1\\
*\\
*
\end{array}\right)
\to
\left(
\begin{array}{c}
\infty^1\\
*\\
*\\
\infty^1
\end{array}\right)
\to
$$

$$
\to
\left(
\begin{array}{c}
*\\
*\\
\infty^1\\
*
\end{array}\right)
\to
\left(
\begin{array}{c}
*\\
\infty^1\\
*\\
-1
\end{array}\right)
\to
\left(
\begin{array}{c}
\infty^1\\
*\\
-1\\
*
\end{array}\right)
\to
\left(
\begin{array}{c}
*\\
-1\\
*\\
*
\end{array}\right)
\to
\left(
\begin{array}{c}
*\\
*\\
*\\
*
\end{array}\right).
$$

The orbit of (let us say) $x_2$ is 
$$*\to (-1)\to *\to \infty^1\to *\to *\to\infty^1\to *\to (-1)\to *.$$

The case of $x_4=-1$ also leads to a confining pattern:
$$
\cdots{\rm regular}
\to
\left(
\begin{array}{c}
a_1\\
a_2\\
a_3\\
-1
\end{array}\right)
\to
\left(
\begin{array}{c}
*\\
*\\
-1\\
\infty^2
\end{array}\right)
\to
\left(
\begin{array}{c}
*\\
-1\\
\infty^2\\
-1
\end{array}\right)
\to
\left(
\begin{array}{c}
-1\\
\infty^2\\
-1\\
*
\end{array}\right)
\to
\left(
\begin{array}{c}
\infty^2\\
-1\\
*\\
*
\end{array}\right)
\to
\left(
\begin{array}{c}
-1\\
*\\
*\\
*
\end{array}\right).
$$

and the orbit of $x_2$ is
$$*\to (-1)\to \infty^2\to (-1)\to *.$$

There is also the possibility of entering through $x_4=0$, which does not produce infinities, but only blow down of subvarities. In this case we also have the strictly confining pattern:

$$
\left(
\begin{array}{c}
a_1\\
a_2\\
a_3\\
0^1
\end{array}\right)
\to
\left(
\begin{array}{c}
*\\
*\\
0^1\\
0^1
\end{array}\right)
\to
\left(
\begin{array}{c}
*\\
0^1\\
0^1\\
*
\end{array}\right)
\to
\left(
\begin{array}{c}
0^1\\
0^1\\
*\\
*
\end{array}\right)
\to
\left(
\begin{array}{c}
0^1\\
*\\
*\\
*
\end{array}\right).
$$

Finally, we have the following confining singularity patterns:
$$ * \to 0 \to 0 \to * ,$$

$$ * \to (-1) \to \infty^2 \to (-1) \to * ,$$

$$ * \to (-1) \to * \to \infty^1 \to * \to * \to \infty^1 \to *\to (-1)\to * .$$

The express method can be applied now. The first one gives us nothing (is the problem of small patterns \cite{mase-1}). The second one yields the following equation:
$$1+\lambda^2-2\lambda=0, \quad |\lambda|_{\rm max}=1.$$
The third one gives
$$1+\lambda^7-\lambda^2-\lambda^5=0,\quad |\lambda|_{\rm max}=1,$$ 
in perfect agreement with integrability. The bilinear substitution involves two tau functions and the resulting equation is strongly multilinear. Unfortunately we could not find any tractable bilinear form.

%In order to find the bilinar form we have to write some symmetry relations:
%We have the so called schwartzian Volterra system:
%$$U_t=\frac{(U_1-U)(U-U_{-1})(U_2-U_{-2})}{(U_2-U_{-1})(U_1-U_{-2})}$$
%If we put $U=G/F$ we obtain:
%$$D_t G\cdot F=\frac{(G_1F-GF_1)(GF_{-1}-G_{-1}F)(G_2F_{-2}-G_{-2}F_2)}{(G_2F_{-1}-G_{-1}F_2)(G_1F_{-2}-G_{-2}F_1)}$$
%which can be bilinearised immediately
%$$D_tG\cdot F=G_2F_{-2}-G_{-2}F_2$$
%$$G_1F-GF_1=G_2F_{-1}-G_{-1}F_2$$

%If we consider 
%$$v=\frac{U-U_1}{U_2-U_{-1}}$$
%then one can see immediately that if $U=G/F$ then by the above bilinear form we find $v=F_2F_{-1}/FF_1$.
%In the same spirit Adler \cite{adler2} found that for SK3 we have the following relation with schwartzian Volterra
%$$x=-\frac{(U_1-U_{-1})(U-U_{-2})}{(U_1-U_{-2})(U-U_{-1})}=\frac{(G_1F_{-1}-G_{-1}F_1)(GF_{-2}-G_{-2}F)}{(G_1F_{-2}-G_{-2}F_1)(GF_{-1}-G_{-1}F)}$$
%Accordingly the bilinear substitution is (showing exactly the $\infty^2$ at the denominator as in the singularity pattern)
%$$x=-\frac{(G_1F_{-1}-G_{-1}F_1)(GF_{-2}-G_{-2}F)}{(GF_{-1}-G_{-1}F)^2}$$
%and the bilinear form:

%\begin{equation}
%D_tG\cdot F=G_2F_{-2}-G_{-2}F_2
%\end{equation}
%\begin{equation}
%G_1F-GF_1=G_2F_{-1}-G_{-1}F_2
%\end{equation}

\section{Conclusions}
The main conclusion that can be drawn is that in higher order differential-difference equation many tau-functions can appear from various singularity patterns. It is not clear which one is the ``good'' one needed for Hirota bilinear form. Apparently (and we saw this in \cite{fane}) the simplest pattern gives the good tau function. However, in the examples analyzed here, all the confining patterns have the same complexity. We managed to find the relations between these tau functions and we constructed the bilinear forms. The problem of different patterns producing {\it many} bilinear equations and the proliferation of tau functions (as in the example (\ref{volt2})) is open. There are also numerous differential-difference equations which may have extremely rich singularity patterns, such as the M\"obius invariant systems \cite{adler2}, \c Ti\c teica and Kaup-Kuperschmidt \cite{adler1,adler2} equations and various variants of Bogoyavlenski lattices \cite{garifullin}. We hope to tackle the relation between singularities and bilinear structure in future publications.
\vspace{0.4 cm}

\textbf{Data availability:}

No data was used for the research described in the article.

\textbf{Conflict of Interests:}

There is no financial and non-financial conflict of interests

\end{document}